\documentclass{aa}

\usepackage{graphicx}

\usepackage{graphicx}

\newcommand{\swift}{{\it Swift}}
\newcommand{\xmm}{{\it XMM-Newton}}

\newcommand{\kepler}{{\it Kepler}}

\newcommand{\ginga}{{\it Ginga}}

\begin{document}

\title{Alternation of the flickering morphology between the high and low state in MV\,Lyr}
\titlerunning{Flickering morphology in high and low state in MV\,Lyr}
\authorrunning{Dobrotka et al.}

\author{A.~Dobrotka \inst {1}, H. Negoro \inst {2} and P. Konopka \inst {3}}

\offprints{A.~Dobrotka, \email{andrej.dobrotka@stuba.sk}}

\institute{Advanced Technologies Research Institute, Faculty of Materials Science and Technology in Trnava, Slovak University of Technology in Bratislava, Bottova 25, 917 24 Trnava, Slovakia
\and
			Department of Physics, Nihon University, 1-8 Kanda-Surugadai, Chiyoda-ku, Tokyo 101-8308, Japan
\and
			Komensk\'eho 4530/6, 92101 Pie\v{s}\v{t}any, Slovakia
}

\date{Received / Accepted}

\abstract
{}
{We studied unique data of a nova-like system MV\,Lyr during transition from the high to low state and vice versa taken by the \kepler\ space telescope. We were interested in evolution of frequency components found previously by Scaringi et al. in different data also obtained by \kepler.}
{We divided the light curve into 10 day segments and investigated the corresponding power density spectra. We searched for individual frequency components by fitting with Lorentzian functions. Additionally, we investigated the variability using averaged shot profiles calculated from the light curve divided into 10 equally spaces subsamples.}
{We found very complex changes of the power density spectra. We focused our study onto three frequency components. Strong activity increase is seen at low frequencies. Contrariwise, the high frequency part of the spectrum strongly decreases in power with specific rise in characteristic frequencies of the individual components. We discuss various scenarios of this phenomenology as reprocessing of X-rays in a receding accretion disc or a radiation from a more active region at the outer disc. Finally, we show that various cataclysmic variables show similar characteristic frequencies in their power density spectra. These are dependent on activity stage, making the situation similar to X-ray binaries.}
{}


\maketitle

\section{Introduction}
\label{introduction}

A great variety of objects such as cataclysmic variables (CVs) are powered by accretion process. It is based on mass loss from a companion star. The transported gas falls towards the central compact object being a white dwarf, and in the absence of a strong magnetic field, an accretion disc forms. The same model is valid also for related objects as X-ray binaries (see e.g. \citealt{warner1995} or \citealt{frank1992} for a review). The main difference is that the latter has a neutron star or a black hole instead of the white dwarf.

The family of CVs is divided into several subclasses based on characteristic variability patterns. Dwarf novae show quasiregular outbursts with durations of several days and appearing on a time scale of 10 - 100 days (see \citealt{warner1995} for review). Nova-like or VY\,Scl systems spend most of their life time in a high state while they sporadically exhibit transition into a low brightness state. This behaviour can be connected to changes in accretion rate. In this way, the duration of the high states in VY\,Scl systems and dwarf novae is significantly different. It is relatively stable and long lasting for the former, while it is only temporary for the latter (outbursts). The shorter durations of the dwarf novae high state can be explained by the limited amount of gas in the depletive disc driving the outburst. The latter is a consequence of a mass accretion rate being unstable because of hydrogen ionization. This phenomenon is well explained by the so called disc instability model (see \citealt{lasota2001} for review). Meanwhile, in VY\,Scl systems, the mass accretion rate remains above the critical limit required for stability, explaining the longer duration of high states. The low state is generated by a sudden drop/stop of mass transfer from the secondary (\citealt{king1998}, \citealt{hessman2000}).

The existence of the accretion process is usually seen as fast stochastic variability (a.k.a. flickering). Flickering has three very basic observational characteristics; 1) linear correlation between variability amplitude and log-normally distributed flux (so called rms-flux relation) observed in all variety of accreting systems (see e.g. \citealt{scaringi2012b}, \citealt{vandesande2015} for CVs case), 2) character of red noise or band limited noise with characteristic frequencies in power density spectra (PDS, see e.g. \citealt{scaringi2012a}, \citealt{dobrotka2014}, \citealt{dobrotka2016} for CVs case) and 3) time lags where flares reach their maxima slightly earlier in the blue than in the red (\citealt{scaringi2013}, \citealt{bruch2015})

An alternative method to "see" the PDS components was proposed by \citet{negoro1994}. The authors superposed many flares (shots) from the \ginga\ light curve of the X-ray binary Cyg\,X-1 in order to get a mean profile showing all typical stable feature like the central spike and two humps on both sides of the spike. A very similar profile was found by \citet{sasada2017} using \kepler\ data of the blazar W2R\,1926+42 but on a much longer time scale. The third object showing such a multicomponent flare shape with a central spike and side-lobes is MV\,Lyr (\citealt{dobrotka2019}) with a time scale between the X-ray binary and the blazar. All three findings make the situation very interesting because of very different nature of all three objects. Especially the radiation origin in the blazar is very different from the other two objects.

MV\,Lyr is a well studied bright nova-like system ideal for flickering study consisting of two phases, i.e. before and after the \kepler\ space telescope. During the former, several studies suggested various variability components. Besides coherent frequencies close to log($f$/Hz) = -4.08 \citet{borisov1992} and \citet{skillman1995} reported a possible quasi periodic oscillation (QPO) at approximately log($f$/Hz) = -3.45. \citet{kraicheva1999} also found a QPO at a similar frequency. First detection of the linear rms-flux relation was presented by \citet{boeva2011}. The latter got an unambiguous shape using extensive \kepler\ data in the study of \citet{scaringi2012b}. \citet{scaringi2012a} used the same \kepler\ data to study the PDS morphology. The authors reported a multicomponent nature with four different characteristic frequencies. Some of these components were close to the previously detected signals. Time lags study in different bands was performed by \citet{scaringi2013} using ULTRACAM on the William Herschel Telescope. For MV\,Lyr 3\,s lags are observed at the lowest frequencies with redder bands lagging the bluer ones. The authors suggest reprocessing of the X-ray photons on to the accretion disc or inside-out shocks traveling within the disc as explanation for the observed lags. \citet{scaringi2014} presented a sandwich model as responsible for the highest PDS component in which the central geometrically thin disc is surrounded by a geometrically thick disc (hot corona). This model was confirmed by direct X-ray observations by \citet{dobrotka2017}.

The localization of the individual PDS components is not trivial and various attempts have been made. Besides the modeling of \citet{scaringi2014}, a shot noise model of \citet{dobrotka2015a} yields very similar results to the former, i.e. the radius of the highest frequency component emission region of approximately $10^{10}$\,cm, and high $\alpha$ parameter (\citealt{shakura1973}) value (close to 1). Moreover, supposing $\alpha$ values from 0.1 to 0.4 the two lower PDS components were localized into the geometrically thin disc and a more active outer disc rim.

The shot noise model is too strong simplification of the real physics, and the localization of the PDS components except the highest one based on \citet{scaringi2014} and \citet{dobrotka2017} is still not certain. Therefore, observational studies are very useful, mainly during dynamic stages of the accretion disc. \kepler\ data of MV\,Lyr comprise such transition to/from the low state typical for nova-like systems (probably) caused by disc reformation. Therefore, we can follow the disc regression and subsequent reconstruction together with PDS evolution during such a high/low state transition. This can yield different flickering source identification and better understanding of the accretion flow geometry or structure of the accretion disc. In this paper we perform detailed analysis of the mentioned activity stage transition of MV\,Lyr observed by the \kepler\ satellite.

\section{Data}

\begin{figure}
\resizebox{\hsize}{!}{\includegraphics[angle=-90]{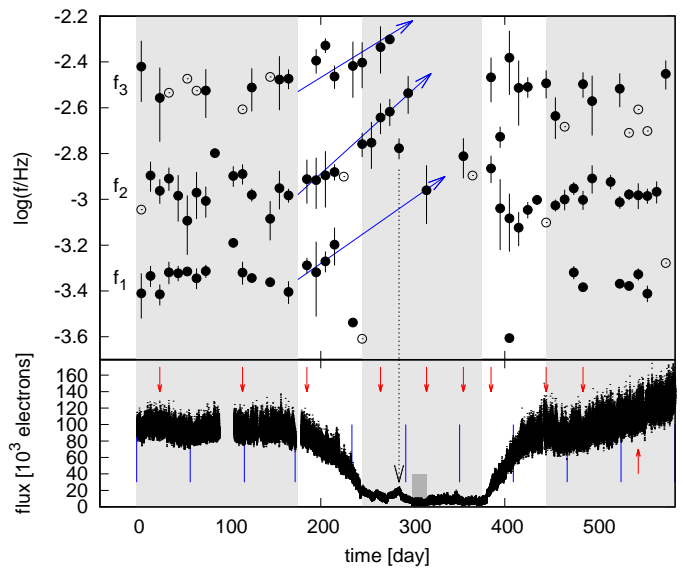}}
\caption{Upper panel - the (time) evolution of PDS components derived using Lorentzians in Eq.~(\ref{equation_pds_model}). The characteristic frequencies $f_{\rm i}$ (marked as labels on the left side) are shown by the black circles with error bars ($\delta {\rm log}(f_{\rm i})$) for significantly detected components with $\delta {\rm log}(f_{\rm i}) < 0.2$, and by black circles without error bars for less significantly detected ones with 0.2 <  $\delta {\rm log}(f_{\rm i})$ < 0.5. The light shaded areas show time intervals selected as before, during and after the low state. Blue arrows outlines possible frequency evolution, and dashed vertical black arrow shows the connection of one deviated point with local re-brightening (see text for details). Lower panel - the analyzed light curve for comparison. The red arrows show the (middle) time location of the PDS examples in Fig.~\ref{pds_evol}. The dark shaded area is the deep low state identified by \citet{scaringi2017} (see Section~\ref{section_shot_profile} for details). The vertical blue lines divide the light curve into 10 equally spaced subsamples for shot profile calculation in Section~\ref{section_shot_profile}.}
\label{pds_lc}
\end{figure}

The data analyzed in this work represent the evolution of MV\,Lyr from the most common high state to a minimum and back. The data are taken by the \kepler\ satellite (\citealt{borucki2010}) with a cadence of approximately 60\,s. The studied light curve lasts approximately 585 days. \kepler\ data comprise relatively many rare null point which we removed. No other correction was needed. The bottom panel of Fig.~\ref{pds_lc} depicts this light curve divided into three intervals marked as shaded areas, i.e. before, during and after the low state.

\section{PDS analysis}

\subsection{Method}
\label{method}

Based on our motivation to study the PDS evolution from the high to low state (and vice versa), we divided the light curve from Fig.~\ref{pds_lc} into equally spaced segments with duration of 10 days. Every segment is divided into ten subsamples, and a periodogram is calculated from each subsample using the Lomb-Scargle algorithm\footnote{\kepler\ data are full of gaps, rare null points and interval of these null points. Therefore, the Lomb-Scargle method is suitable because it is designed for such unevenly spaced data. Otherwise discrete Fourier transform would be ideal.} \citet{scargle1982}. All periodograms (power\footnote{We use power normalized by the total variance according to \citet{horne1986}. Since the normalization does not affect the shape of the PDS, this has no importance in our case.} $p$ as function of frequency $f$) were subsequently transformed into log($f$)-log($p$) space, because of several reasons. Firstly, \citet{papadakis1993} concluded that for examination of variations of the PDS with time (to detect QPOs), the periodogram averaging in logarithm is better rather than averaging the periodograms themselves. Secondly, such averaging of log($p$) yields symmetric errors (see e.g. \citealt{vanderklis1989}, \citealt{aranzana2018}). Thirdly, the whole log($f$) interval of the PDS is re-binned with a constant frequency step of 0.05, and such frequency bins comprise more and more periodogram points toward higher frequencies. The PDS scatter rises toward higher frequencies when binned with a constant frequency step in a linear scale (Fig.~4 of \citealt{scaringi2012a}), and any potential frequency component above log($f$/Hz) = -3 is totally buried and non detectable. Finally, all log($p$) points within each frequency bin were averaged\footnote{If the number of averaged log-log periodogram points per bin is lower than the selected value of 30 (3 points from each of 10 periodograms), the bin is larger until the condition is fulfilled. This happens for lowest frequencies. The motivation is to get enough points for mean value with error determination.}, with the standard error of the mean as an uncertainty estimate. Such PDS estimate has equal resolution in the whole log($f$) interval (see e.g. \citealt{shahbaz2005}, \citealt{aranzana2018}).

The periodogram averaging and binning lower the PDS scatter, while the light curve division determines the frequency resolution being equal to the lowest PDS frequency (before re-binning). The latter is proportional to the shortest light curve subsample duration. Therefore, an empirical compromise between noise and resolution must be found. The high frequency PDS end is set up empirically. Usually it is equal to the frequency where Poisson noise becomes dominant so that the PDS becomes flat (white noise) with no additional information. We concentrated our study to the high frequency part of the PDS, i.e. we excluded the two lowest (from total of four) PDS components detected by \citet{scaringi2012a} because of low frequency resolution of the PDS. The two remaining high frequency components were detected also in X-rays by \citet{dobrotka2017} facilitating any discussion and interpretation. We also searched for an additional high frequency component which is of X-ray origin too (\citealt{dobrotka2019}).

\subsection{Fitted model}
\label{fits}

\begin{figure}
\resizebox{\hsize}{!}{\includegraphics[angle=-90]{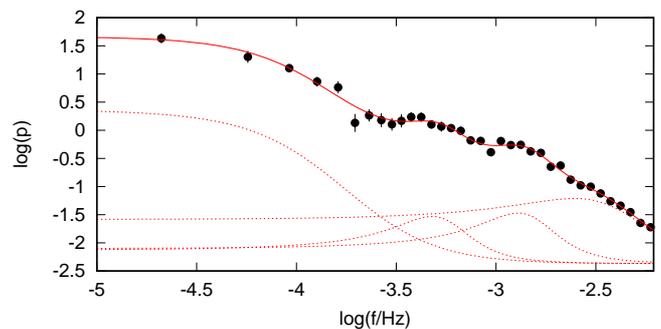}}
\caption{Example of individual Lorentzian components (red dashed line) from Equation~(\ref{equation_pds_model}) yielding a multicomponent shape of the fit (red solid line). While the fitting is performed in log($p$) vs $f$, the individual PDS "humps" are better seen in log($f \times p$) vs log($f$) (see such equivalent in Fig.~\ref{pds_evol}, day 110-120).}
\label{pds_comp_ex}
\end{figure}

Every PDS is fitted with a multicomponent model using {\small GNUPLOT}\footnote{http://www.gnuplot.info/}. We used the same model as \citet{scaringi2012a}, but applied it in a slightly different way. The model consists of $n$ Lorentzian functions taken from \citet{belloni2002}
\begin{equation}
p = \sum_{i=1}^{n} \left[ \frac{c_{\rm i} \Delta_{\rm i}}{\pi}\frac{1}{\Delta_{\rm i}^2 + (f - f_{\rm i})^2} \right],
\label{equation_pds_model}
\end{equation}
where $c_{\rm i}$ is a constants, $f_{\rm i}$ is the searched characteristic frequency of the corresponding PDS components and $\Delta_{\rm i}$ is its half-width at half-maximum. All PDSs were fitted with $n = 4$ if possible. If any component was not identified visually, or the fitting process did not yield a satisfying result, we reduced the number of components up to $n = 2$. Usually the fourth (lowest) Lorentzian fitting the lowest frequency part converged to a very low value of $f_{\rm i}$ and described the rising or constant power toward lowest frequencies. This component is not of our interests and is used just to describe continuously the whole PDS (Fig.~\ref{pds_comp_ex}).

Finally, when fitting the PDSs we used the averaged log($p$) instead of $p$ (described above). All subsequent PDSs are better visualized in ${\rm log}(f \times p)$ as y-axis units. Such visualization is suitable in the case of steep red noise, where individual components are better seen which simplifies the characteristic frequencies identification needed for the number of components set up and initial parameter estimate.

\subsection{Results}
\label{results}

\begin{figure}
\resizebox{\hsize}{!}{\includegraphics[angle=-90]{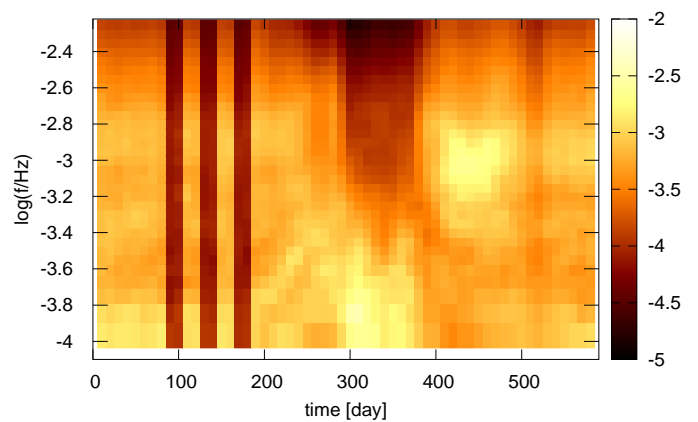}}
\caption{PDS evolution as a function of time with y-axis units of log$(f \times p)$. The color/gray scale represents the normalized Lomb-Scargle power. The three vertical dark bands are gaps due to too large gaps in the data. The map is slightly smoothed.}
\label{map_pds}
\end{figure}

\begin{figure}
\resizebox{\hsize}{!}{\includegraphics[angle=-90]{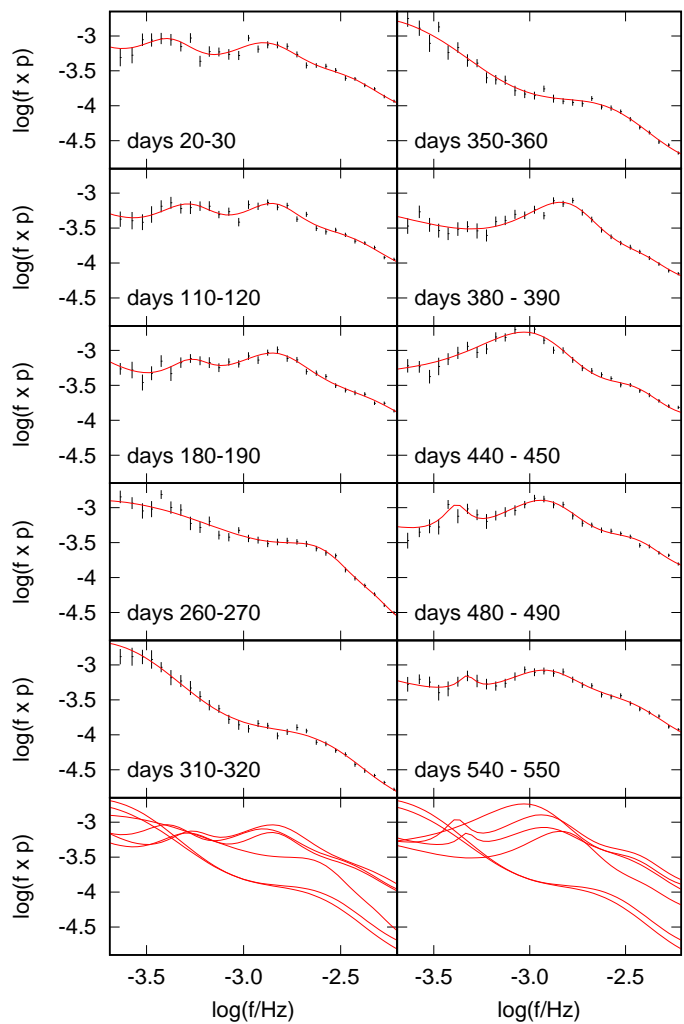}}
\caption{Some representative examples of the calculated PDSs (small vertical lines as errors of the mean) with corresponding fits (red lines) in $f \times p$ as y-axis units. Two different cases are shown, i.e. transitions from the high to low state (left panels) and vice versa (right panels). The (middle) time locations of the PDSs are marked as red arrows in the lower panel of Fig.~\ref{pds_lc}. All displayed fits are directly compared in the bottom panels (both panels comprise days 310-320 and 350-360 for comparison).}
\label{pds_evol}
\end{figure}

The power of individual frequencies is depicted as an evolution map in Fig.~\ref{map_pds}, with some examples shown in Fig.~\ref{pds_evol}. The decrease of the high frequency power during the low state is clear, while the power rises below approximately log($f$/Hz) = -3.4. Moreover, another strong power increase at approximately log($f$/Hz) = -3 after the minimum is noticeable too.

For further analysis and discussion of the PDS evolution we selected only the best and well resolved fits of clear PDS components, i.e. we selected PDS frequencies with errors $\delta {\rm log}(f_{\rm i})$ < 0.2. All bad or poor fits yield large errors due to large PDS data scatter. Individual frequencies $f_{\rm i}$ are depicted in upper panel of Fig.~\ref{pds_lc}.

Three discrete frequencies before and after the minimum are noticeable. While these values seem to be stable during the high state, the frequencies are variable during the transition. During the latter the $f_1$ component is missing. As seen in Fig.~\ref{pds_evol} the PDS dramatically changes in shape in this dynamic stage, i.e. it becomes steep with hardly resolvable components. Therefore, it is not sure whether the components disappeared or their powers just become too small to be resolved. Moreover, two low frequency points at log($f$/Hz) = -3.54 and -3.61 (days 235 and 405, respectively) appears during the transitions. Whether these belong to $f_1$ or they are independent components is hard to conclude because of a small number of fitted frequencies. Helpful is the evolution map in Fig.~\ref{map_pds} where a clear line is seen to rise from log($f$/Hz) = -3.8 to -3.6 between days 200 and 300. This suggests, that the two deviated low frequency points represent an independent component with frequency lower than $f_1$.

\begin{figure}
\resizebox{\hsize}{!}{\includegraphics[angle=-90]{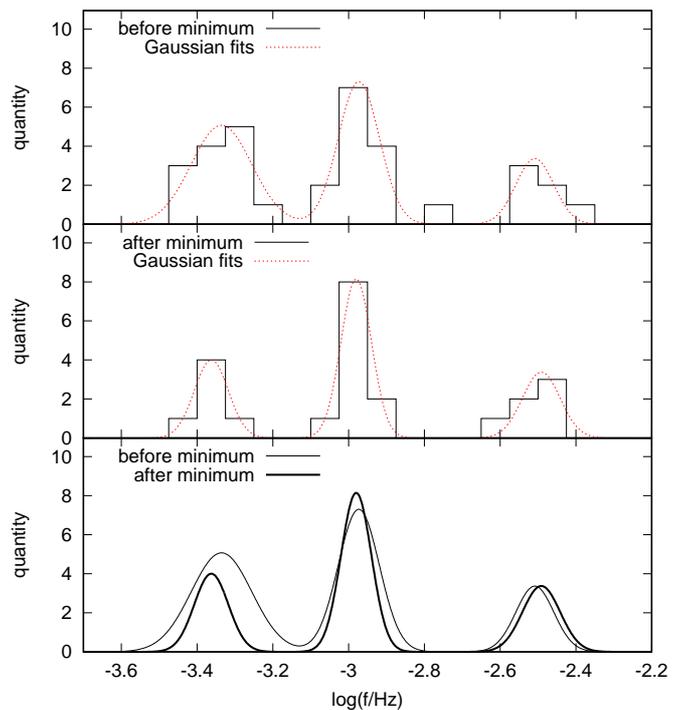}}
\caption{Histograms of measured frequencies from the upper panel of Fig.~\ref{pds_lc} (points in the shaded areas are used). Multi-components Gaussian fits are shown and compared in the bottom panel.}
\label{hist}
\end{figure}

The distribution of frequencies before and after the transition is well described by histograms in Fig.~\ref{hist}. We fit individual cases with a multi-Gaussian model with three components. The fitted mean values with 1-$\sigma$ as errors are summarized in Table~\ref{pds_param}. Direct comparison of the fits (bottom panel of Fig.~\ref{hist}) suggests that after the minimum all frequencies gain their pre-minimum values.
\begin{table}
\caption{Mean values with 1-$\sigma$ parameters as error estimates of the frequencies from Fig.~\ref{pds_lc} before and after the low state (marked as gray areas in Fig.~\ref{pds_lc}).}
\begin{center}
\begin{tabular}{lccc}
\hline
\hline
position & log($f_1$/Hz) & log($f_2$/Hz) & log($f_3$/Hz)\\
\hline
before & $-3.34 \pm 0.08$ & $-2.97 \pm 0.05$ & $-2.51 \pm 0.05$\\
after & $-3.36 \pm 0.05$ & $-2.98 \pm 0.04$ & $-2.49 \pm 0.05$\\
\hline
\end{tabular}
\end{center}
\label{pds_param}
\end{table}

We did not do the same histogram study for the low state because the points suggest variable behaviour. Just at the beginning of the brightness fall all three components start to deviate toward higher frequencies suggesting systematic variability (marked as blue arrows in Fig.~\ref{pds_lc}). The trend continues toward day 300 after which the detection of individual PDS components was very problematic or even impossible.

There are two interesting points during the low state in Fig.~\ref{pds_lc}. One is at day 290. It does not follow the increasing trend of $f_2$, but jumps bellow other measurements. This point corresponds to a local re-brightening marked as vertical dashed arrow. Probably this opposite brightness behaviour corresponds to opposite frequency evolution. The other deviated point is during the day 320. It appears like pre-minimum $f_2$ value, but this can be also $f_1$ component which evolved to such a high value (at the end of the corresponding blue arrow). After the low state, all frequencies return back to the pre-minimum level.

\section{Superposed shot profile analysis}
\label{section_shot_profile}

The PDS is a very generalized technique, and various structures with the same/similar characteristic frequency can blend together and form a single PDS feature. This is the case of the dominant PDS feature at log($f$/Hz) = -3. \citet{dobrotka2019} performed a superposed shot profile analysis (method based on \citealt{negoro1994}) of the fast variability, and found that both the central spike and side lobes produce a PDS pattern very close to log($f$/Hz) = -3. If this PDS component is variable, the standard PDS study can not resolve substructures of which the shot profile is responsible for the variability.

\subsection{Method}

Therefore, as an additional study of the variability morphology we performed the same superposed shot profile analysis as in \citet{dobrotka2019}. The goal is to "see directly" individual PDS components in the flare shape like in the case of Cyg\,X-1 (\citealt{negoro2001}). The method has three steps. First is the shots/flares identification. A light curve point is identified as a peak, if $N_{\rm pts}$ points to the left and $N_{\rm pts}$ points to the right have lower fluxes than the tested point.

As a second step, the flare extension must be defined, i.e. $N_{\rm ptsext}$ points to the left and to the right from the peak. Like in \citet{dobrotka2019} we used $N_{\rm ptsext} = N_{\rm pts}/2$ in order to not superimpose a declining branch of one flare with a rising branch of the adjacent flare, and vice versa.

As a last step after the flare selection, the flare points are averaged with maxima aligned, and resulting averaged flux minimum is subtracted from all averaged points. All flares with rare individual null points or missing data (at the edge of the light curve or gaps) were excluded from the averaging process. As discussed in \citet{dobrotka2019} the long-term trend and barycentric correction of \kepler\ data do not affect the result.

As $N_{\rm pts}$ we used a value of 50. This results in 51 points (with a duration of approximately 58.8\,s) per flare yielding a flare duration of 3060\,s with a corresponding frequency of log($f$/Hz) = -3.49. This is enough to visualize the studied frequency interval in the previous section.

For a superposed shot profile we used 10 equally spaced subsamples of the whole light curve as shown by the vertical blue lines in Fig.~\ref{pds_lc}. This results in enough flares for averaging and we can visualize the rough time evolution. A larger number of light curve subsamples would yield to a smaller number of flares per average resulting in higher noise and lower resolution.

Finally, \citet{scaringi2017} investigated a part of the same data (during the low state) and found a quasi-periodic signal present only during the very faintest period of time when the light curve reaches a roughly constant minimum brightness level. The authors call this state a deep low state. They present a scenario in which this non-magnetic CVs undergoes a short lasting regime of magnetically driven accretion causing the quasi-periodic bursts. In order to not contaminate the average profile with other burst-like patterns, we excluded the deep low state from the analysis. This data interval between days 299 and 315 is marked as dark shaded area in the lower panel of Fig.~\ref{pds_lc}. The QPOs have a shape of 30\,min bursts repeating every 2\,hours. The corresponding frequencies are log($f$/Hz) = -3.86 and -3.25. The upper panel of Fig.~\ref{pds_lc} does not show any point at these frequencies during the problematic part, therefore the PDS is not affected.

\subsection{Results}

\begin{figure}
\resizebox{\hsize}{!}{\includegraphics[angle=-90]{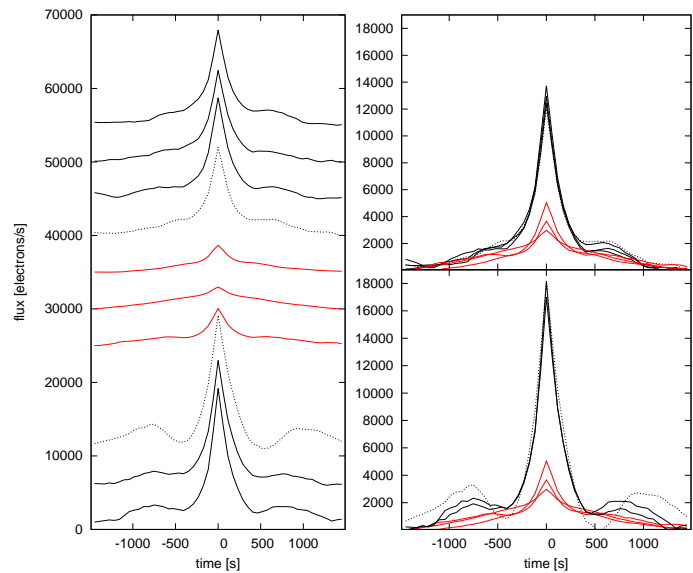}}
\caption{Averaged shot profile time evolution. Black solid lines represent shot profiles during the high state, the black dotted lines are the profiles during both transitions, and the red solid lines represent the low state. Left panel - profiles vertically offset with time elapsing from up to down. Right panels - direct comparison/superposition of profiles showing accession to the low state (upper panel), and ascension from the low state (lower panel).}
\label{shot_evol}
\end{figure}

Resulting shot profiles with a central spike and side-lobes like in \citet{dobrotka2019} are shown in Fig.~\ref{shot_evol}. During the transition from the high to low state (upper right panel) all components amplitudes just decrease, with side-lobes being non-detectable at the end. During the transition from the low to high state (lower right panel) the behaviour is reverse, and the central spike gets higher amplitude compared to the pre-minimum stage. While its amplitude increases and remains more or less stable after the transition, the amplitude of the side-lobes culminates toward the end of the transition phase (dotted line). The latter probably corresponds to the enhanced power of the $f_2$ component at the end of the transition phase seen in Fig.~\ref{map_pds} or in Fig.~\ref{pds_evol} as days 440-450. This is the information not provided by the standard PDS method, i.e. different shot profile substructures with the same/similar characteristic frequency behaving differently blends into a single PDS component.

\section{Discussion}
\label{discussion}

We studied the PDS evolution of MV Lyr before, during and after the low state captured by the \kepler\ satellite. We focused our study on frequencies detected in previous PDS analysis by \citet{scaringi2012a}. We used a slightly different method than the authors in order to reduce the PDS scatter. We averaged/binned the periodograms in log-log space with a constant logarithmic frequency step if a minimum number of averaged periodogram points per bin is satisfied. The latter reduces the resolution at low frequencies. Therefore, we were interested only in PDS components with highest frequencies.

\subsection{Phenomenology}

All PDS components are probably present in both the high and low states with variable values during the transition and low state. The two highest components show significant increase in frequency. The lowest frequency component is problematic to detect in the low state, but if present, it increases in frequency too. The power above log($f$/Hz) = -3.4 strongly decreases during the transition to the low state, while it significantly increases for lower frequencies. This different behaviour suggests different physical origin of the variability below and above log($f$/Hz) = -3.4.

The characteristic frequency variability during the transition suggests the correlation of the frequencies with the flux. However, after the system returns back into the high state, the flux rises continuously while the frequency values are stable. Therefore, the frequencies depend on the activity stage and not the flux. This may be expected, because if the disc is fully reconstructed after the low state, no additional structural changes are generated by the increasing mass accretion rate. The idea is that the characteristic frequencies depend on the accretion flow structures (reconstructed and stable after the minimum), while the flux is proportional to the mass accretion rate (still rising after the minimum).

\subsection{Physical model}

\subsection{Varying inner disc radius}

The presence of PDS components in various accreting systems motivates many authors to connect the frequency values with physical parameters of that system. For example the connection of black hole mass with a PDS break frequency in active galactic nuclei is such case (see e.g. \citealt{gonzales2012}, \citealt{mohan2014}). This should be the result of different inner disc radii being a function of the black hole mass. In CVs the PDS components can be connected to the inner disc radius too (\citealt{balman2012}). The luminosity of these binaries is dominated by the release of gravitational potential energy of the gas in the disc, and therefore the brightness is directly connected to the mass accretion rate. The sudden drop in brightness in MV\,Lyr or VY\,Scl binaries (see \citealt{warner1995} for a review) means that the mass accretion rate decreases considerably. The mass transfer can fall from more than $10^{-8}$\,M$_{\rm \odot}$\,y$^{-1}$ to less than $10^{-11}$\,M$_{\rm \odot}$\,y$^{-1}$ (\citealt{scaringi2017}). The latter is typical for a dwarf novae in quiescence. Following the disc instability model (see \citealt{lasota2001} for a review) the disc is truncated during such a low state, while it is fully reformed down to the white dwarf surface in the high state. Therefore, during the transition from the high to low state studied in this paper, the inner disc must recede to larger radii. This is the basic idea of \citet{balman2012} who suggest that the characteristic frequency of a PDS component generated at the inner disc radius must decrease simultaneously. However, the variability of all components in this paper shows an opposite behaviour. This suggests that the mentioned correlation of the characteristic frequencies with the inner disc radius is not applicable.

\subsection{$f_2$, $f_3$ components and reprocessing model}

\begin{figure}
\resizebox{\hsize}{!}{\includegraphics[angle=0]{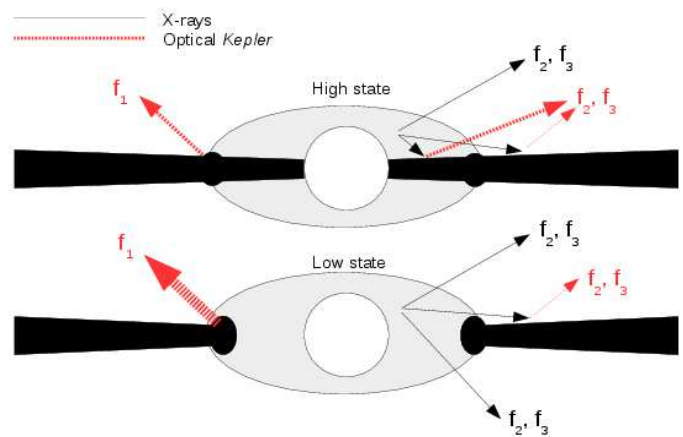}}
\caption{Visualization of the proposed scenario, where $f_2$ and $f_3$ components are generated in the hot corona as X-ray radiation which is reprocessed into optical by the underlying thin disc. The thickness of the arrows represents the radiation intensity. See text for details.}
\label{model}
\end{figure}

Another explanation of the observed behaviour is based on the model proposed by \citet{scaringi2014}. Following the author the $f_2$ component is generated by an inner geometrically thick disc, the so called X-ray corona. This disc radiates in X-rays, which are subsequently reprocessed into optical by the inner geometrically thin disc. \citet{dobrotka2017} confirmed the X-ray origin of this PDS component, which supports the corona origin. Moreover, the $f_3$ component is of X-ray origin too (\citealt{dobrotka2019}). If the thin disc is reprocessing the X-rays into optical, the recession of this disc during the low state means lower reprocessing surface. This can result in optical flux decrease while the frequencies are unchanged (Fig.~\ref{model}). The variation of the latter can be generated by the transforming corona due to changing mass accretion rate.

Another aspect is the X-ray emissivity. During the high state only a small fraction of the flow is evaporated into the corona. The mass accretion rate based on a cooling flow model yields only $7 \times 10^{-12}$\,M$_{\rm \odot}$\,y$^{-1}$ (\citealt{dobrotka2017}). This is typical value of the mass accretion rate in the low-state of this class of objects (see e.g. \citealt{gansicke1999}), when the central disc is truncated and the whole mass flow is evaporated into the corona (Fig.~\ref{model}). The X-rays are generated by free-free transitions and the emissivity is proportional to $n^2$, where $n$ is the plasma density. If the mass accretion rate (or $n$) through the corona is the same (unchanged) during both the high and low states, the gravitational energy liberation does not vary. This results in unaltered amplitude of the variability, and no PDS power decrease due to emissivity is expected. Apparently, this is not observed as shown in this paper. Therefore, additional physical process (ex. reprocessing) affecting the variability is expected.

Finally, the temporal higher power of the $f_2$ component after the minimum is interesting. During the disc redevelopment the propagating increase of the mass accretion from outer disc regions pushes the inner disc edge toward the white dwarf. Such a mass wave can enhance specific variability in the corona, which returns to the initial pre-minimum stage after relaxation/stabilization of the flow.

\subsection{Energetic aspect of the reprocessed X-rays}

The reprocessing scenario is attractive, but some energy based test is worth to do. For this purpose we used \xmm\ OM data presented in \citet{dobrotka2017} because these data are well calibrated, while the \kepler\ case is more problematic. After transforming the OM light curve taken with UVW1 filter into fluxes, the amplitude of the variability represented by the root square of the variance is of $2.3 \times 10^{-14}$\,ergs/s/cm/\AA. Assuming the same flux in the whole optical band we get a rough estimate of the integrated optical flux. However, this is merely an upper limit because it is well known that the flickering has higher amplitude toward higher energies. So in reality, the amplitude should decrease toward lower energies. Lets suppose that the wavelength interval of our assumed "optical" band is between 2500 and 7000\,\AA\ (limits of UVW1\footnote{Using lower energy filters like U or B starting at 3000 or 3500\,\AA\ instead of UVW1 does not change the following energetic discussion.} and V filter). Multiplying the UVW1 flux with the length of this band (4500\,\AA) we get the integrated flux of $1.4 \times 10^{-10}$\,ergs/s/cm, i.e. due to roughness of the estimate we take $\sim 10^{-10}$\,ergs/s/cm. Moreover, \citet{dobrotka2017} mention that the \swift\ flux in the interval of 0.2 - 10\,keV is approximately $5.4 \times 10^{-12}$\,ergs/s/cm, i.e. approximately 20 times less than the estimated optical flux upper limit. This makes the reprocessing scenario energetically insufficient.

We do not expect that the radiation outside the 0.2 - 10\,keV range contributes significantly to the overall X-ray flux. Therefore, the estimated X-ray flux is at its maximum. In order to get both energy bands closer, more realistic estimate of the optical flux is the only possibility. Some reduction of the latter can be achieved when accounting for decreasing flux toward lower energies. If we assume that the flickering amplitude decreases linearly, and reaches zero value at the lower energy limit of the V filter\footnote{This allows some amplitude to be observable also in the R band.}, the reduction factor is 2. This does not resolve the energy insufficiency, it only decreases the band ratio from 20 to 10.

Another aspect is the time scale of the flickering variability. \citet{scaringi2014} modeled only the $f_2$ frequency, and identified the corona as the variability source. Therefore, in such case the reprocessing generates only the optical $f_2$ signal, and the lower frequency components can be generated by the accretion disc itself. Therefore, in order to get correct energy estimate, the lower frequencies must be excluded from the flickering. For this purpose we detrended the OM light curve from Fig.~1 in \citet{dobrotka2017} using running median. Using 20 adjacent points\footnote{10 points to the left, 10 to the right from the corrected point. This yields time scale of 1050\,s (including the corrected point) with the OM light curve cadence of 50\,s, which corresponds approximately to the frequency $f_2$.} for the running window from which the median is calculated, the $f_2$ feature is still dominant in the PDS. The corresponding amplitude of the variability is approximately 1.5 times lower. Such reduction of the variability amplitude is again not a solution, and it only decreases the band ration from 10 to approximately 7. A lower number of points per median window already affects the dominant $f_2$ PDS feature, and therefore any further reduction of the variability amplitude is not possible.

\begin{figure}
\resizebox{\hsize}{!}{\includegraphics[angle=-90]{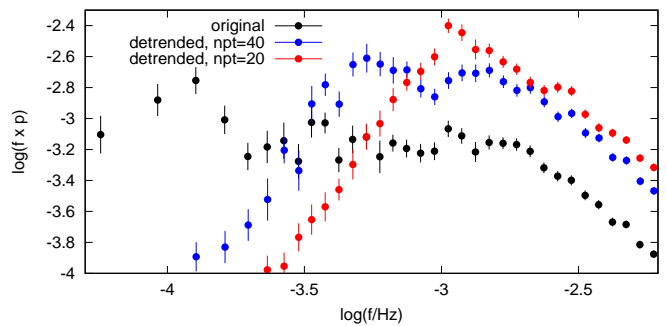}}
\caption{PDSs from the first 10 days of \kepler\ data. Compared are the original case with two PDSs calculated from data after detrending using running median. $npt$ represents the number of median window points.}
\label{pds_detrend}
\end{figure}

\begin{figure}
\resizebox{\hsize}{!}{\includegraphics[angle=-90]{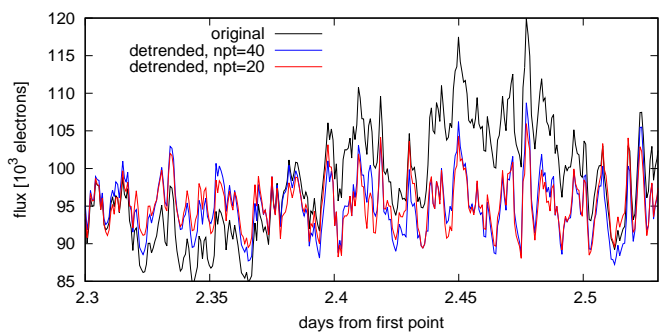}}
\caption{Snapshot of the \kepler\ light curve compared with the detrended cases after removing the running median. The low frequency variability not present in the red and blue curves is clear. Parameter $npt$ has the same meaning as in Fig.~\ref{pds_detrend}.}
\label{lc_detrend}
\end{figure}

The \kepler\ case is negligibly better. In Fig.~\ref{pds_detrend} we show the original and detrended PDSs of first 10 days. Detrending using 40 and 20 points per median window reduces the variability amplitude by a factor of approximately 1.8 and 2, respectively (Fig.~\ref{lc_detrend}). Apparently, all "attempts" how to reduce the optical flux are insufficient, and all estimates are still one order of magnitude larger than the estimated X-ray flux.

Apparently, we have a serious inconsistency between interpretation and energies. The clue is in an exact understanding of the variability components. Relying only on PDS, we study one $f_2$ pattern. But the shot profile studied in Section~\ref{section_shot_profile} and in \citet{dobrotka2019} identify two $f_2$ components which blend and are non-distinguishable in standard PDS study. We propose a model, where the inner disc mass accretion fluctuations are somehow correlated with the fluctuations in the corona. The inner disc generates its own high amplitude $f_2$ pattern, while the corona generates a low amplitude $f_2$ pattern by reprocessing. Such high amplitude pattern can be the central spike seen in Fig.~\ref{shot_evol}, while the low amplitude patterns can be the side lobes. Inspecting the amplitudes from Fig.~5 in \citet{dobrotka2019} we find that the side lobe amplitudes are lower than the peak one by factors of 8 and 9.2 for the left and right side lobe, respectively. Taking the profiles from Fig.~\ref{shot_evol}, the factors are comparable except the one with clearly enhanced side lobes. This reduces the corresponding optical flux by one order of magnitude, making the energetic ratio satisfied and suitable for the reprocessing scenario.

Finally, the corona is an evaporation of the underlying thin disc, and any mass accretion fluctuation of this thin disc must results in modulation of the mass evaporation. Such fluctuations in evaporation can modulate and trigger mass accretion fluctuations in the corona. This is the above assumed disc-corona correlation.

\subsection{Low frequency components}

The behaviour of the $f_1$ component during the low state is enigmatic because of poor statistics. However, strong power rise toward lower frequencies is noticeable. Whether it represents the shifting $f_1$ is not sure, but certainly it suggests increasing activity below $f_1$. Not reacting to inner disc recession like $f_2$ and $f_3$ suggests that the origin of this power rise must be searched elsewhere. Some outer disc region with possible enhanced activity during the low state must be a source. For example the disc-corona interaction (Fig.~\ref{model}) at the corona outer edge can be a solution.

Such a more active region can be a source of QPOs generated by rotational motion of accretion inhomogeneity. This rotation would generate variabilities on a dynamical time scale
\begin{equation}
t_{\rm dyn} \sim \left( \frac{R^3}{G\,M_1} \right)^{1/2},
\end{equation}
where G is the gravitational constant, $M_1$ the primary mass and $R$ the distance form the center. Using $M_1$ = 0.73\,M$_{\rm \odot}$ (\citealt{hoard2004}) $t_{\rm dyn} = 72$ or 580\,s for $R = 8 \times 10^9$ and $32 \times 10^9$\,cm, respectively. The former is the corona radius derived by \citet{scaringi2014}, while the latter is a disc radius estimate from \citet{dobrotka2015a}. Corresponding frequencies of log($f$/Hz) $\sim -2.8$ and -1.9 are too high for $f_1$ or lower values. However, the value of -2.8 is close to log($f_2$/Hz) $\sim -3.0$ component in the high state. But from \citet{scaringi2014} and \citet{dobrotka2017} we know that viscous process in the geometrically thick X-ray corona is the probable source in this case.

\begin{figure}
\resizebox{\hsize}{!}{\includegraphics[angle=-90]{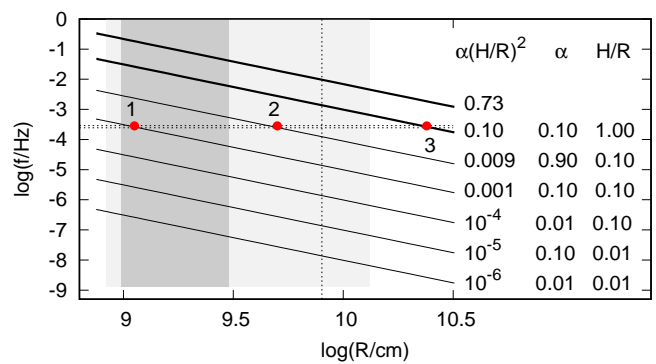}}
\caption{Frequencies corresponding to viscous time scale compared to the two lowest measured frequencies from Fig.~\ref{pds_lc} (two horizontal dotted lines). Various models with different parameters listed as labels are shown. The vertical dotted line is the corona radius from \citet{scaringi2014}. The shaded areas are the inner disc radius intervals from Fig.~4 of \citet{scaringi2017}. Labeled points are solutions showing the measured frequencies.}
\label{t_visc}
\end{figure}

Since we study the disc reconfiguration driven by viscous processes, similar estimate can be done for the viscous time scale
\begin{equation}
t_{\rm visc} \sim \frac{t_{\rm dyn}}{\alpha (H/R)^2},
\end{equation}
where $\alpha$ is the basic disc parameter by \citet{shakura1973} and $H$ is the height scale of the disc. Some examples of the corresponding frequencies as a function of $R$ are shown in Fig.~\ref{t_visc}. The values are calculated from the white dwarf surface estimated following \citet{nauenberg1972} up to the outer disc edge of $32 \times 10^9$\,cm. The vertical dotted line shows the corona radius from \citet{scaringi2014}. Various combinations of $\alpha$ and $H/R$ are shown. Apparently, compared to the horizontal dotted lines showing the frequencies of log($f_1$/Hz) = -3.54 and -3.61 (lowest values from Fig.~\ref{pds_lc}), large range of $\alpha (H/R)^2$ is possible, i.e. from 0.001 to 0.1 shown as 3 solutions (points 1, 2, 3) yielding $H/R \geq 0.1$. This suggests that the source is no longer a geometrically thin disc. The solutions range from (point 1) an inner radius disc derived by \citet{scaringi2017} using maximum possible constrains (dark shaded area in Fig.~\ref{t_visc}) up to (point 3) a radius larger than the coronal radius derived by \citealt{scaringi2014}. Therefore, we are not able to quantify the source characteristics, only conclusion is that the geometrically thick corona is a potential solution.

Other disc structures can be still candidates for the low frequency components. A hot spot or stream (from the secondary) overflow are also localized further away from the central disc. The superhump interpretation would be also possible, but the discussed frequencies log($f$/Hz) = -3.54 and -3.61 are far from the superhump case of log($f$/Hz) = -4.08 derived by \citet{borisov1992} and \citet{skillman1995}. Only direct X-ray observation can bring final answer. Crucial is to determine whether $f_1$ is of X-ray origin like $f_2$ and $f_3$, and how is the PDS power compared to the high state. If detected in X-rays, this would exclude any stream overflow, hot spot or superhump scenario, and places the radiation source into central regions like a corona or boundary layer.

\subsection{Analogy with X-ray binaries}

The presence of PDS components in \kepler\ data of CVs at frequencies close to the values determined in this paper is not new. \citet{dobrotka2015} and \citet{dobrotka2016} studied PDSs of two dwarf novae present in the \kepler\ field, i.e. V1504\,Cyg and V344\,Lyr. Both systems show similar behaviour of the PDSs in dependence of the activity stage. Surprisingly not only the frequency values are comparable, but also their presence depending of the activity stage. During the low state both systems show PDS components close to $f_1$, while during the high state another components close to $f_2$ are present.

Some other systems studied from the ground show similar low frequency components, i.e. KR\,Aur (\citealt{kato2002}) and UU\,Aqr (\citealt{baptista2008}). While the former shows practically the same characteristic PDS frequency as derived in this paper, the latter is slightly lower\footnote{\citealt{baptista2008} reported a presence of spiral arm and this can significantly change the nature of the variability, i.e. it may differ from the "standard" case.}. \citet{balman2012} reported several high frequency components close to $f_3$ in X-ray data of SS\,Cyg, RU\,Peg, VW\,Hyi, WW\,Cet and T\,Leo. \citet{dobrotka2014} reanalyzed the observation of the dwarf nova RU\,Peg during quiescence and found the multicomponent nature of the PDS with similar characteristic frequencies in UV and X-rays. Contrary to V1504\,Cyg and V344\,Lyr, the quiescent X-ray PDS also comprises a component between $f_2$ and $f_3$. Perhaps more or all components are present in both stages of CVs in general, but some remains very damped during the low state. After all, the $f_2$ and $f_3$ components in MV\,Lyr did not disappear during the low state, just the power is very low. Because of the latter and probable variability the characteristic values are hard to be determined.

\begin{figure}
\resizebox{\hsize}{!}{\includegraphics[angle=-90]{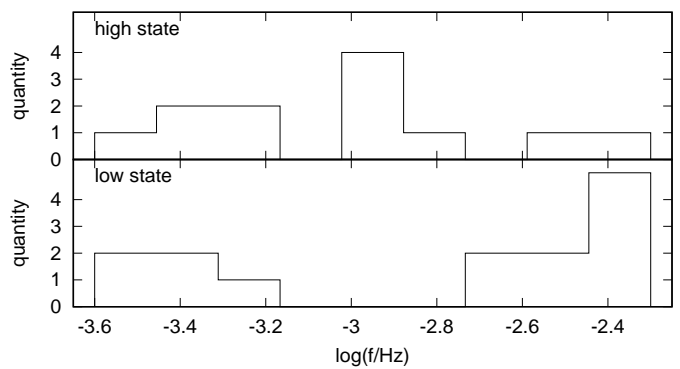}}
\caption{Histograms of the values from Table~\ref{pds_param_compar}. The value of -3.8 was excluded because of spiral patterns in the disc of UU\,Aqr making the case "non standard".}
\label{pds_cvs}
\end{figure}

All the mentioned values are summarized in Table~\ref{pds_param_compar} and Fig.~\ref{pds_cvs}. Three distinct groups are potentially present. However, confirmation of this statement requires much larger statistical set of measured frequencies, because the histogram in Fig.~\ref{pds_cvs} can be a result of a random process.
\begin{table}
\caption{List of approximate characteristic PDS frequencies in different CVs systems used for Fig.~\ref{pds_cvs}. The values in parenthesis are maximal (not necessary final) values of $f_2$ and $f_3$ from Fig.~\ref{pds_lc}. We did not add the hypothetical highest value of $f_1$ (in the low state) because it is too speculative.}
\begin{center}
\begin{tabular}{lccl}
\hline
\hline
system & state & band & log($f$/Hz)\\
\hline
MV\,Lyr$^a$ & low & optical & -3.5, (-2.5)\\
 & low & optical & -3.6, (-2.3)\\
V1504\,Cyg$^b$ & low & optical & -3.4, -2.3\\
V344\,Lyr$^c$ & low & optical & -3.4 \\
RU\,Peg$^d$ & low & X-rays & -3.5, -2.7\\
& low & UV & -3.3\\
SS\,Cyg$^e$ & low & X-rays & -2.3\\
& low & X-rays & -2.4\\
VW\,Hyi$^e$ & low & X-rays & -2.7\\
WW\,Cet$^e$ & low & X-rays & -2.5\\
T\,Leo$^e$ & low & X-rays & -2.4\\
\hline
MV\,Lyr$^a$ & high & optical & -3.3, -3.0, -2.5\\
& & & -3.4\\
V1504\,Cyg$^b$ & high & optical & -3.3, -3.0, -2.9, -2.4\\
V344\,Lyr$^c$ & high & optical & -3.5, -3.0, -2.8\\
KR\,Aur$^f$ & high & optical & -3.4\\
UU\,Aqr$^g$ & high & optical & -3.8\\
\hline
\end{tabular}
\end{center}
$^a$ this paper\\
$^b$ from Table~2 of \citet{dobrotka2015}\\
$^c$ from Table~2 of \citet{dobrotka2016}\\
$^d$ from \citet{dobrotka2014}\\
$^e$ from \citet{balman2012}\\
$^f$ from \citet{kato2002}\\
$^g$ from \citet{baptista2008}
\label{pds_param_compar}
\end{table}

Such typical PDS components depending on brightness states are very typical for X-ray binaries (see e.g. \citealt{miyamoto1992}, \citealt{miyamoto1993}, \citealt{miyamoto1994} as pioneer works, or \citealt{lewin2010} for a review), while it is new for CVs.

\section{Summary and conclusions}

The multicomponent PDS of MV Lyr shows complex transformation during the transition from high to low state and vice versa with three resolved PDS components. The high frequency part of the PDS (log($f$/Hz) > -3.4) considerably decreases in power during the minimum brightness compared to the high state, while the low frequency part (log($f$/Hz) < -3.4) significantly increases. The characteristic frequencies of the two highest PDS components increase during the low state. The same is also possible for the lowest PDS component, but with lower significance. Once the brightness starts to get back to the pre-minimum level, the frequencies returns to their pre-minimum values. Additional transient power increase of the middle PDS component at log($f$/Hz) $\simeq$ -3 is clear after the minimum. The evolution of frequencies together with the overall flux suggests that the frequency values are correlated with the brightness state, and not with the flux.

The power decrease of the high frequency part of the PDS can be explained by diminishing of the receding disc surface needed for X-ray reprocessing. This is based on the knowledge that the highest PDS components are generated by the inner X-ray corona or boundary layer. However, rough estimates of the integrated optical and X-ray fluxes imply that the reprocessing scenario can not work for the whole variability, but only for its low amplitude components. Rough time scale estimate implies that the source of the lowest frequencies can be a standard disc (geometrically thicker than thin solution) embedded into the geometrically thick X-ray corona or the corona itself.

Finally, the comparison with other CV systems suggests that CVs may have common typical PDS components with similar values depending on the brightness state. This is similar to what is observed in X-ray binaries. However, larger statistical set of measurements is needed to confirm or decline this hypothesis.

\section*{Acknowledgment}

AD was supported by the Slovak grant VEGA 1/0408/20, and by the Operational Programme Research and Innovation for the project : Scientific and Research Centre of Excellence SlovakION for Material and Interdisciplinary Research“, code of the project ITMS2014+ : 313011W085 co-financed by the European Regional Development Fund. HN was supported by Grants-in-Aid for Scientific Research 16K05301 from the Ministry of Education, Culture, Sports, Science and Technology (MEXT) of Japan. We acknowledge the anonymous referee for helpful comments, particularly concerning the energetic study of the optical and X-ray flickerings.

\bibliographystyle{aa}
\bibliography{mybib}

\begin{thebibliography}{46}
\expandafter\ifx\csname natexlab\endcsname\relax\def\natexlab#1{#1}\fi

\bibitem[{{Aranzana} {et~al.}(2018){Aranzana}, {K{\"o}rding}, {Uttley},
  {Scaringi}, \& {Bloemen}}]{aranzana2018}
{Aranzana}, E., {K{\"o}rding}, E., {Uttley}, P., {Scaringi}, S., \& {Bloemen},
  S. 2018, \mnras, 476, 2501

\bibitem[{{Balman} \& {Revnivtsev}(2012)}]{balman2012}
{Balman}, {\c S}. \& {Revnivtsev}, M. 2012, \aap, 546, A112

\bibitem[{{Baptista} \& {Bortoletto}(2008)}]{baptista2008}
{Baptista}, R. \& {Bortoletto}, A. 2008, \apj, 676, 1240

\bibitem[{{Belloni} {et~al.}(2002){Belloni}, {Psaltis}, \& {van der
  Klis}}]{belloni2002}
{Belloni}, T., {Psaltis}, D., \& {van der Klis}, M. 2002, \apj, 572, 392

\bibitem[{{Boeva} {et~al.}(2011){Boeva}, {Bachev}, {Tsvetkova}, {Stoyanov},
  {Zamanov}, {Spassov}, {Latev}, {Petrov}, {Donchev}, {Dimitrov}, {Valcheva},
  \& {Georgiev}}]{boeva2011}
{Boeva}, S., {Bachev}, R., {Tsvetkova}, S., {et~al.} 2011, Bulgarian
  Astronomical Journal, 16, 23

\bibitem[{{Borisov}(1992)}]{borisov1992}
{Borisov}, G.~V. 1992, \aap, 261, 154

\bibitem[{{Borucki} {et~al.}(2010)}]{borucki2010}
{Borucki}, W.~J. {et~al.} 2010, Science, 327, 977

\bibitem[{{Bruch}(2015)}]{bruch2015}
{Bruch}, A. 2015, \aap, 579, A50

\bibitem[{{Dobrotka} {et~al.}(2014){Dobrotka}, {Mineshige}, \&
  {Ness}}]{dobrotka2014}
{Dobrotka}, A., {Mineshige}, S., \& {Ness}, J.-U. 2014, \mnras, 438, 1714

\bibitem[{{Dobrotka} {et~al.}(2015){Dobrotka}, {Mineshige}, \&
  {Ness}}]{dobrotka2015a}
{Dobrotka}, A., {Mineshige}, S., \& {Ness}, J.-U. 2015, \mnras, 447, 3162

\bibitem[{{Dobrotka} {et~al.}(2019){Dobrotka}, {Negoro}, \&
  {Mineshige}}]{dobrotka2019}
{Dobrotka}, A., {Negoro}, H., \& {Mineshige}, S. 2019, \aap, 631, A134

\bibitem[{{Dobrotka} \& {Ness}(2015)}]{dobrotka2015}
{Dobrotka}, A. \& {Ness}, J.-U. 2015, \mnras, 451, 2851

\bibitem[{{Dobrotka} {et~al.}(2016){Dobrotka}, {Ness}, \& {Baj{\v c}i{\v
  c}{\'a}kov{\'a}}}]{dobrotka2016}
{Dobrotka}, A., {Ness}, J.-U., \& {Baj{\v c}i{\v c}{\'a}kov{\'a}}, I. 2016,
  \mnras, 460, 458

\bibitem[{{Dobrotka} {et~al.}(2017){Dobrotka}, {Ness}, {Mineshige}, \&
  {Nucita}}]{dobrotka2017}
{Dobrotka}, A., {Ness}, J.-U., {Mineshige}, S., \& {Nucita}, A.~A. 2017,
  \mnras, 468, 1183

\bibitem[{{Frank} {et~al.}(1992){Frank}, {King}, \& {Raine}}]{frank1992}
{Frank}, J., {King}, A., \& {Raine}, D. 1992, Cambridge Astrophysics Series, 21

\bibitem[{{G{\"a}nsicke} {et~al.}(1999){G{\"a}nsicke}, {Sion}, {Beuermann},
  {Fabian}, {Cheng}, \& {Krautter}}]{gansicke1999}
{G{\"a}nsicke}, B.~T., {Sion}, E.~M., {Beuermann}, K., {et~al.} 1999, \aap,
  347, 178

\bibitem[{{Gonz{\'a}lez-Mart{\'{\i}}n} \& {Vaughan}(2012)}]{gonzales2012}
{Gonz{\'a}lez-Mart{\'{\i}}n}, O. \& {Vaughan}, S. 2012, \aap, 544, A80

\bibitem[{{Hessman}(2000)}]{hessman2000}
{Hessman}, F.~V. 2000, \nar, 44, 155

\bibitem[{{Hoard} {et~al.}(2004){Hoard}, {Linnell}, {Szkody}, {Fried}, {Sion},
  {Hubeny}, \& {Wolfe}}]{hoard2004}
{Hoard}, D.~W., {Linnell}, A.~P., {Szkody}, P., {et~al.} 2004, \apj, 604, 346

\bibitem[{{Horne} \& {Baliunas}(1986)}]{horne1986}
{Horne}, J.~H. \& {Baliunas}, S.~L. 1986, \apj, 302, 757

\bibitem[{{Kato} {et~al.}(2002){Kato}, {Ishioka}, \& {Uemura}}]{kato2002}
{Kato}, T., {Ishioka}, R., \& {Uemura}, M. 2002, \pasj, 54, 1033

\bibitem[{{King} \& {Cannizzo}(1998)}]{king1998}
{King}, A.~R. \& {Cannizzo}, J.~K. 1998, \apj, 499, 348

\bibitem[{{Kraicheva} {et~al.}(1999){Kraicheva}, {Stanishev}, \&
  {Genkov}}]{kraicheva1999}
{Kraicheva}, Z., {Stanishev}, V., \& {Genkov}, V. 1999, \aaps, 134, 263

\bibitem[{{Lasota}(2001)}]{lasota2001}
{Lasota}, J. 2001, \nar, 45, 449

\bibitem[{{Lewin} \& {van der Klis}(2010)}]{lewin2010}
{Lewin}, W. \& {van der Klis}, M. 2010, {Compact Stellar X-ray Sources}

\bibitem[{{Miyamoto} {et~al.}(1993){Miyamoto}, {Iga}, {Kitamoto}, \&
  {Kamado}}]{miyamoto1993}
{Miyamoto}, S., {Iga}, S., {Kitamoto}, S., \& {Kamado}, Y. 1993, \apjl, 403,
  L39

\bibitem[{{Miyamoto} {et~al.}(1994){Miyamoto}, {Kitamoto}, {Iga}, {Hayashida},
  \& {Terada}}]{miyamoto1994}
{Miyamoto}, S., {Kitamoto}, S., {Iga}, S., {Hayashida}, K., \& {Terada}, K.
  1994, \apj, 435, 398

\bibitem[{{Miyamoto} {et~al.}(1992){Miyamoto}, {Kitamoto}, {Iga}, {Negoro}, \&
  {Terada}}]{miyamoto1992}
{Miyamoto}, S., {Kitamoto}, S., {Iga}, S., {Negoro}, H., \& {Terada}, K. 1992,
  \apjl, 391, L21

\bibitem[{{Mohan} \& {Mangalam}(2014)}]{mohan2014}
{Mohan}, P. \& {Mangalam}, A. 2014, \apj, 791, 74

\bibitem[{{Nauenberg}(1972)}]{nauenberg1972}
{Nauenberg}, M. 1972, \apj, 175, 417

\bibitem[{{Negoro} {et~al.}(2001){Negoro}, {Kitamoto}, \&
  {Mineshige}}]{negoro2001}
{Negoro}, H., {Kitamoto}, S., \& {Mineshige}, S. 2001, \apj, 554, 528

\bibitem[{{Negoro} {et~al.}(1994){Negoro}, {Miyamoto}, \&
  {Kitamoto}}]{negoro1994}
{Negoro}, H., {Miyamoto}, S., \& {Kitamoto}, S. 1994, \apjl, 423, L127

\bibitem[{{Papadakis} \& {Lawrence}(1993)}]{papadakis1993}
{Papadakis}, I.~E. \& {Lawrence}, A. 1993, \mnras, 261, 612

\bibitem[{{Sasada} {et~al.}(2017){Sasada}, {Mineshige}, {Yamada}, \&
  {Negoro}}]{sasada2017}
{Sasada}, M., {Mineshige}, S., {Yamada}, S., \& {Negoro}, H. 2017, \pasj, 69,
  15

\bibitem[{{Scargle}(1982)}]{scargle1982}
{Scargle}, J.~D. 1982, \apj, 263, 835

\bibitem[{{Scaringi}(2014)}]{scaringi2014}
{Scaringi}, S. 2014, \mnras, 438, 1233

\bibitem[{{Scaringi} {et~al.}(2013){Scaringi}, {K{\"o}rding}, {Groot},
  {Uttley}, {Marsh}, {Knigge}, {Maccarone}, \& {Dhillon}}]{scaringi2013}
{Scaringi}, S., {K{\"o}rding}, E., {Groot}, P.~J., {et~al.} 2013, \mnras, 431,
  2535

\bibitem[{{Scaringi} {et~al.}(2012{\natexlab{a}}){Scaringi}, {K{\"o}rding},
  {Uttley}, {Groot}, {Knigge}, {Still}, \& {Jonker}}]{scaringi2012a}
{Scaringi}, S., {K{\"o}rding}, E., {Uttley}, P., {et~al.} 2012{\natexlab{a}},
  \mnras, 427, 3396

\bibitem[{{Scaringi} {et~al.}(2012{\natexlab{b}}){Scaringi}, {K{\"o}rding},
  {Uttley}, {Knigge}, {Groot}, \& {Still}}]{scaringi2012b}
{Scaringi}, S., {K{\"o}rding}, E., {Uttley}, P., {et~al.} 2012{\natexlab{b}},
  \mnras, 421, 2854

\bibitem[{{Scaringi} {et~al.}(2017){Scaringi}, {Maccarone}, {D'Angelo},
  {Knigge}, \& {Groot}}]{scaringi2017}
{Scaringi}, S., {Maccarone}, T.~J., {D'Angelo}, C., {Knigge}, C., \& {Groot},
  P.~J. 2017, \nat, 552, 210

\bibitem[{{Shahbaz} {et~al.}(2005){Shahbaz}, {Dhillon}, {Marsh}, {Casares},
  {Zurita}, {Charles}, {Haswell}, \& {Hynes}}]{shahbaz2005}
{Shahbaz}, T., {Dhillon}, V.~S., {Marsh}, T.~R., {et~al.} 2005, \mnras, 362,
  975

\bibitem[{{Shakura} \& {Sunyaev}(1973)}]{shakura1973}
{Shakura}, N.~I. \& {Sunyaev}, R.~A. 1973, \aap, 24, 337

\bibitem[{{Skillman} {et~al.}(1995){Skillman}, {Patterson}, \&
  {Thorstensen}}]{skillman1995}
{Skillman}, D.~R., {Patterson}, J., \& {Thorstensen}, J.~R. 1995, \pasp, 107,
  545

\bibitem[{{Van de Sande} {et~al.}(2015){Van de Sande}, {Scaringi}, \&
  {Knigge}}]{vandesande2015}
{Van de Sande}, M., {Scaringi}, S., \& {Knigge}, C. 2015, \mnras, 448, 2430

\bibitem[{{van der Klis}(1989)}]{vanderklis1989}
{van der Klis}, M. 1989, in NATO Advanced Science Institutes (ASI) Series C,
  Vol. 262, NATO Advanced Science Institutes (ASI) Series C, ed.
  H.~{{\"O}gelman} \& E.~P.~J. {van den Heuvel}, 27

\bibitem[{{Warner}(1995)}]{warner1995}
{Warner}, B. 1995, Cambridge Astrophysics Series, 28

\end{thebibliography}

\label{lastpage}

\end{document}